\pdfoutput=1
\documentclass[a4paper,aps,preprintnumbers,showpacs,twocolumn,superscriptaddress,nofootinbib,amsmath,amssymb,floatfix]{revtex4}
\usepackage[T1]{fontenc}
\usepackage[utf8]{inputenc}
\usepackage{cmap}
\usepackage{graphicx}
\usepackage{multirow,hhline}
\usepackage{ulem}

\newcommand{\ie}{{i.e.,}~}
\newcommand{\eg}{{e.g.,}~}

\begin{document}
\title{Stability of asymptotically de Sitter and anti-de Sitter black holes in $4D$ regularized Einstein-Gauss-Bonnet theory}

\author{M. A. Cuyubamba}
\affiliation{Institute of Physics, University of Oldenburg, D-26111 Oldenburg, Germany}
\email{marcoance1624@gmail.com}

\begin{abstract}
The regularized four-dimensional Einstein-Gauss-Bonnet model has been recently proposed in [D. Glavan and C. Lin, Phys. Rev. Lett. \textbf{124}, 081301 (2020)] whose formulation is different of the Einstein theory, allowing us to bypass the Lovelock theorem. The action is formulated in higher dimensions ($D>4$) by adding the Gauss-Bonnet correction to the conventional Einstein-Hilbert action with a cosmological constant. The four-dimensional spacetime is constructed through dimensional regularization by taking the limit $D\rightarrow 4$. We find explicitly the parametric regions of stability of black holes for the asymptotically flat and (anti-)de Sitter spacetimes by analyzing the time-domain profiles for gravitational perturbations in both vector and scalar channels. In addition to the known eikonal instability we find the instability due to the positive cosmological constant. On the contrary, asymptotically anti-de Sitter black holes have no other instability than the eikonal one.
\end{abstract}
\pacs{04.50.Kd,04.70.Bw,04.25.Nx,04.30.-w}
\maketitle

\section{Introduction}
Alternative theories of gravity have gained interest in recent years \cite{Clifton:2011jh} in attempt to solve the incompatibility between general relativity (GR) and quantum mechanics, which is one of the fundamental questions of the modern physics. A promising approach to add the quantum corrections to the classical gravity is the higher-dimensional ($D>4$) Lovelock theory \cite{Lovelock:1971} which is constructed by adding higher-curvature corrections to the Einstein action. Due to the Lovelock theorem the model is reduced to Einstein theory in $D=4$. In five or six dimensions, there is an additional term in the action, called the Gauss-Bonnet term, which is quadratic in curvature and can be interpreted as the first-order quantum correction appearing in the low energy limit of the heterotic string theory. For higher-dimensional spacetimes, higher order corrections appear.

Stability of black holes against small perturbations is a crucial condition for viability of a black-hole model \cite{Konoplya:2011qq}. Linear black hole perturbations can be decomposed in a superposition of the characteristic oscillations, called quasinormal modes, which can be enumerated by the multiple number $\ell$. When higher growing rate modes appear for larger multiple number $\ell$, the threshold of instability is called the {\it eikonal instability} emphasizing the fact that it happens in the regime of geometrical optics \cite{Cuyubamba:2016cug}. In the general Lovelock theory, the eikonal instabilities of gravitational perturbations exist for sufficiently small black holes \cite{Takahashi:2010ye,Takahashi:2010gz,Yoshida:2015vua,Dotti:2005sq,Gleiser:2005ra,Konoplya:2017lhs,Konoplya:2017zwo}. For the Einstein-Gauss-Bonnet (EGB) black holes in addition to the eikonal instability, which appears in the asymptotically flat case \cite{Konoplya:2008ix,Konoplya:2017ymp}, there is an instability, which develops at lower multipoles when the cosmological constant is sufficiently large, called, therefore, the {\it $\Lambda$-instability} \cite{Cuyubamba:2016cug}. Likewise, stability of four-dimensional black holes with the second-order of curvature (Gauss-Bonnet term) non-minimally coupled to a scalar field (dilaton) has been studied in recent years \cite{Blazquez-Salcedo:2020rhf,Blazquez-Salcedo:2016enn,Zinhailo:2019rwd,Konoplya:2019hml}.

Lovelock's theorem states that the four-dimensional Einstein tensor with the cosmological constant forms the unique combination which is divergence-free, symmetric and second-order in its equation of motion \cite{Lovelock:1972vz}. Recently it has been proposed a way to circumvent the Lovelock theorem and avoid the Ostrogradsky instability \cite{Glavan:2019inb}. Namely, a four-dimensional spacetime is defined as the limit $D\to 4$ of higher-dimensional Lovelock theory, after the re-scaling of the Gauss-Bonnet coupling constant $\alpha_2\to 2\alpha_2/(D-4)$ in the Lagrangian.
A further generalization, which includes the higher-curvature correction due to the Einstein-Lovelock theory, has been proposed in \cite{Casalino:2020kbt,Konoplya:2020qqh}.
Some properties of black holes in the asymptotically flat $4D$ regularized Einstein-Gauss-Bonnet theory, such as shadows, instability, quasinormal modes of scalar and electromagnetic test fields, and gravitational perturbations, have been studied in \cite{Konoplya:2020bxa,Churilova:2020aca}. Thermodynamics of the $4D$ Einstein-Gauss-Bonnet-anti-de Sitter black holes with electric charge was discussed in \cite{Fernandes:2020rpa} and quasinormal modes of the neutral asymptotically AdS black hole were considered in \cite{Aragon:2020qdc}. The influence of the re-scaled coupling parameter upon the deflection of electromagnetic radiation due to the strong gravitational lensing by a static spherically symmetric Einstein-Gauss-Bonnet black hole has been investigated in \cite{Islam:2020xmy}. Properties of the innermost stable circular orbit (ISCO) of massive, spinning and charged test particles around the spherically symmetric black hole have been studied within this novel model in \cite{Guo:2020zmf,Shaymatov:2020yte,Abdujabbarov:2020jla,Zhang:2020qew}. Parametric region of the eikonal instability for $4D$ Einstein-Gauss-Bonnet and Einstein-Lovelock black holes was obtained in \cite{Konoplya:2020juj}. Some further properties, axial symmetry, Hawking radiation and thermodynamics were studied in \cite{Wei:2020ght,Hegde:2020xlv,Konoplya:2020cbv,Devi:2020uac,Konoplya:2020ibi,Zhang:2020qam,Wang:2020pmb}.

Recently, it was pointed out that the regularization approach encounters some problems. It was shown in \cite{Ai:2020peo} that the approach of Glavan and Lin~\cite{Glavan:2019inb} requires embedding of the four-dimensional spacetime into the higher $D$-dimensional one, and the resulting reduced theory gains additional degrees of freedom. It was shown in \cite{Gurses:2020ofy,Gurses:2020rxb} that the field equations in the $4D$ EGB gravity in the limit ($D\to 4$) contain a $D$-dimensional term (which corresponds to the Lanczos-Bach tensor), which vanishes in simple cases, \eg for the spherically symmetric configurations. However, in general case, $D\to4$ leads to a divergence, preventing a covariant formulation of the corresponding equations. A similar analysis has been performed in \cite{Mahapatra:2020rds} at the level of action. It was concluded that the action and surface terms split into two parts, and one of them does not scale $\propto (D-4)$. Therefore, there are divergent terms in the action as $D\to4$.

If we consider a theory, which leads to the same solutions as we obtain within the regularized approach then, according to Lovelock's theorem, such a four-dimensional theory of gravity should either add extra degrees of freedom or break the diffeomorphism invariance. Some theories with the Gauss-Bonnet term coupled to the scalar field have been proposed in \cite{Hennigar:2020lsl,Fernandes:2020nbq,Lu:2020iav,Easson:2020mpq}. These theories are introduced through the Kaluza-Klein-like compactification and are special cases of the Horndeski theory. However, such theories lead to problems due to the infinitely coupled degrees of freedom \cite{Kobayashi:2020wqy}. A consistent description of the $4D$ Gauss-Bonnet theory is given by Atsuki Aoki et. al. \cite{Aoki:2020lig}, using the Hamiltonian formalism. The theory either breaks the temporal diffeomorphism invariance or leads to an additional (scalar) degree of freedom.

In the present paper we find the parametric region of linear stability for black holes in the asymptotically de Sitter and anti-de Sitter spacetimes, by systematically studying perturbation profiles in the time domain. For the stable black holes we also calculate the dominant quasinormal modes of the scalar-type (polar) and vector-type (axial) gravitational perturbations.

The paper is organized as follows. In Sec. II we present a general framework of a spherically symmetric, static black hole solution within the $4D$ regularized Einstein-Gauss-Bonnet model and its gravitational perturbations in the scalar and vector sectors. In Sec II we present the numerical method, which was used to obtain the temporal profiles for the gravitational perturbations of the black hole. In section IV we are discussing the parametric region of stability for the $4D$ regularized Einstein-Gauss-Bonnet-(anti)de Sitter black holes. Finally, in the conclusion we summarize the results and give a brief outlook of open problems.

\section{Theoretical framework}
\subsection{Static black hole solution}\label{ft}
The Lovelock theory formulated in \cite{Lovelock:1971}, is given by the Lagrangian density
\begin{equation}\label{lovelock-lagrangian}
  \mathcal{L}=-2\Lambda+\sum_{m=1}^{\bar{m}}\frac{\alpha_m}{m}\mathcal{L}_m,
\end{equation}
where $\Lambda$ is the cosmological constant, $\mathcal{L}_m$ are the Lovelock terms and $\alpha_m$ are arbitrary parameters of the theory, such that $2m<D$. For $D=4$ we consider $\bar{m}=1$ in the summation, for $D=5$ or $6$ we have also the second order in curvature, $\bar{m}=2$. Higher than six dimensions can also contain higher-order Lovelock terms. Following \cite{Konoplya:2020qqh}, we define the parameters $\tilde{\alpha}_m$ such as
\begin{equation}\label{alpha-regular}
  \tilde{\alpha}_m=\frac{\alpha_m}{m}\prod_{k=1}^{2m-2}(D-2-k).
\end{equation}
In the limit $D\to 4$, the Gauss-Bonnet coupling constant is
\begin{equation}\label{alpha2-regular}
  \alpha_2\to\frac{2\tilde{\alpha}_2}{(D-4)},
\end{equation}
so that in order to regularize the four dimensional Einstein-Lovelock theory, we shall consider finite value of $\tilde{\alpha}_2$. Such approach has been proposed in \cite{Glavan:2019inb} although initially suggested by Y.~Tomozawa in~\cite{Tomozawa:2011gp}. We limit our consideration by the second order in curvature, \ie consider only the Gauss-Bonnet correction, then the Lagrangian density can be written as
\begin{eqnarray}
  \mathcal{L}&=&-2\Lambda+R+\alpha_2\,\mathcal{L}_2\label{EGB-Lagrangian}\\\nonumber
  &=&-2\Lambda+R+\alpha_2\left(R_{\mu\nu\lambda\sigma}R^{\mu\nu\lambda\sigma}-4R_{\mu\nu}R^{\mu\nu}+R^2\right).
\end{eqnarray}
In general, the coupling constant $\alpha_2$ is a real parameter bounded by some values of the theory. Despite $D\to 4$ has no Lagragian formalism, (the Lagrangian diverges), the second order equations for the metric and consequently the static black-hole solution are well defined, and the metric can be described by the following line element,
\begin{eqnarray}
  ds^2&=&-f(r)dt^2+\frac{1}{f(r)}dr^2+r^2\left(d\theta^2+\sin^2\theta\,d\phi^2\right),\nonumber\\
  f(r)&=&1-r^2\psi(r).\label{metric}
\end{eqnarray}
The radial function $\psi(r)$ satisfies the following relation,
\begin{equation}\label{Ppsi-equation}
  P[\psi(r)]\equiv\psi(r)+\tilde{\alpha}_2\,\psi(r)^2=\frac{2M}{r^3}+\frac{\Lambda}{3},
\end{equation}
which could be obtained after substituting the metric (\ref{metric}) \cite{Boulware:1985wk} into the vacuum Einstein-Gauss-Bonnet equation and considering the limit $D\to4$. The constant of integration $M$ is the asymptotic mass. Furthermore, from (\ref{Ppsi-equation}) we can see that the solution has two branches, and one of them is perturbative in $\alpha_2$,
\begin{equation}\label{function4DEGB}
  f(r)=1-\frac{r^2}{2\tilde{\alpha}_2}\left(-1+\sqrt{1+4\tilde{\alpha}_2\left(\frac{2M}{r^3}+\frac{\Lambda}{3}\right)}\right),
\end{equation}
while the other solution diverges as $\alpha_2\to 0$. Measuring all quantities in units of the event horizon radius $r_H$, we find the mass
\begin{equation}\label{massM}
  2M=r_H\left(1+\frac{\tilde{\alpha}_2}{r_H^2}-\frac{\Lambda r_H^2}{3}\right).
\end{equation}
In the de Sitter spacetime ($\Lambda>0$) with the cosmological horizon $r_C$ ($f(r_C)=0$), the cosmological constant reads as
\begin{equation}\label{lambdarh}
  \frac{\Lambda r_H^2}{3}=\frac{r_H^2}{r_H^2+r_C r_H+r_C^2}\left(1-\frac{\tilde{\alpha}_2}{r_H r_C}\right),
\end{equation}
The extremal value of the cosmological constant is obtained by taking limit $r_C\to r_H$ in (\ref{lambdarh}),
\begin{equation}\label{lambdext}
\Lambda_{ext}=\frac{r_H^2-\tilde{\alpha}_2}{r_H^4}.
\end{equation}

For asymptotically Anti-de Sitter spacetime ($\Lambda<0$), it is convenient to introduce the AdS-radius $R$, which is defined via the asymptotic of the metric function $f(r)\to 1+r^2/R^2$ at infinity. The cosmological constant measure in units of $R$ is
\begin{equation}\label{lambdaR}
  \frac{\Lambda R^2}{3}=-1+\frac{\tilde{\alpha}_2}{R^2}.
\end{equation}

In order to avoid a naked singularity, the function $\psi'(r)$ must be finite outside the event horizon,
\begin{equation}\label{psiprime}
  \psi'(r)=-\frac{6M}{r^3(1+2\tilde{\alpha}_2\psi(r))}.
\end{equation}
Thus, $\tilde{\alpha}_2$ is bounded as
\begin{eqnarray}
\nonumber-0.5r_H^2<&\tilde{\alpha}_2&<r_H^2, \qquad\qquad \mbox{for de Sitter,}\\
\nonumber-0.5r_H^2<&\tilde{\alpha}_2&<0.5 R^2, \qquad \mbox{for anti-de Sitter.}
\end{eqnarray}

\subsection{Gravitational perturbation}
After perturbing the background metric $\delta g_{\mu\nu}$,
\begin{equation}\label{perturb}
  g_{\mu\nu}\rightarrow g_{\mu\nu}+\delta g_{\mu\nu},\qquad\qquad \left|\delta g_{\mu\nu}\right|\ll\left|g_{\mu\nu}\right|,
\end{equation}
the equations which govern the propagation of such perturbation can be reduced to differential wave-like equations,
\begin{equation}\label{wave-like}
  \left(-\frac{\partial^2}{\partial r_*^2}+\frac{\partial^2}{\partial t^2}-V_i(r_*)\right)\Psi(t,r_*)=0
\end{equation}
where $r_*$ is the tortoise coordinate $dr_*=dr/f(r)$ and $V_i$ is the effective potential, where $i$ stands for $t$ (tensor), $v$ (vector) and $s$ (scalar). In the linear approximation, these kinds of perturbations can be treated independently owing to the transformation law of the rotation group on a $(D-2)$-sphere. In the limit $D\to 4$, the tensor-type perturbation becomes purely gauge transformations, possessing no additional degrees of freedom. The effective potentials for the vector- and scalar-type perturbations were determined in \cite{Takahashi:2010ye} and can be expressed as
\begin{eqnarray}
  V_v(r)&=&\frac{(\ell-1)(\ell+2)}{r\,T(r)}\frac{d\,T(r)}{dr_*}+ R(r)\frac{d^2}{dr_*^2}\left(\frac{1}{R(r)}\right),\label{efectivepotentialvector}\\
  V_s(r)&=&\frac{\ell(\ell+1)}{r^2\,B(r)}\frac{d}{dr_*}\left(r\,B(r)\right)+ B(r)\frac{d^2}{dr_*^2}\left(\frac{1}{B(r)}\right),\label{efectivepotentialscalar}
\end{eqnarray}
where $\ell=2,3,4,\ldots$ is the multiple number and
\begin{eqnarray}
  T(r) &=& r\left(1+2\tilde{\alpha}_2\psi(r)\right), \\
  R(r) &=& r\sqrt{T^\prime(r)}, \\
  B(r) &=& \frac{2(\ell-1)(\ell+2)-2r^3\psi^\prime}{R(r)}T(r).
\end{eqnarray}

Notice that the function $T'(r)$ must be positive definite outside the black-hole $r>r_H$ in order to avoid a wrong sign in the kinetic term of the effective potential. The negative kinetic term leads to unbounded growth of initial perturbations which was called ghost-instability \cite{Takahashi:2010ye}.

\section{Characteristic integration}

\begin{figure}
\centering
\resizebox{\linewidth}{!}{\includegraphics*{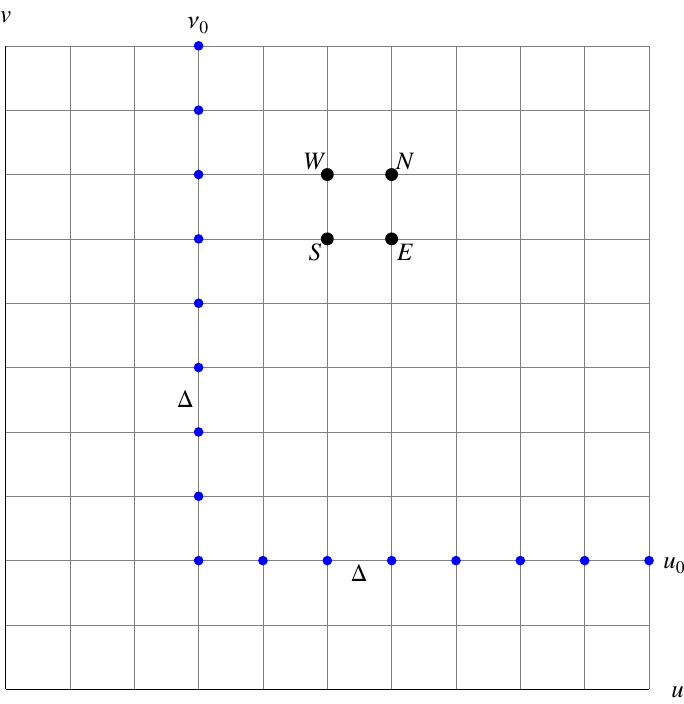}}
\caption{Discretized grid to determine the time evolution of black-hole perturbations. The blue points correspond to the initial perturbations.}\label{profile-03}
\end{figure}

In order to catch the threshold of instability we used the characteristic method proposed by Gundlach, Price, and Pullin in \cite{Gundlach:1993tp}, which allows us to determine the temporal profiles of black holes for any given set of parameters. This method consists in a discretization of a certain region of the spacetime near the black hole, defined by two null surfaces $(u_0,v_0)$ (as we can see in figure \ref{profile-03}) and calculation of the magnitude of the wave-function $\Psi$ on each point of the grid. Rewriting (\ref{wave-like}) in terms of the light-cone coordinates $du=dt-dr^*$ and $dv=dt+dr^*$, one obtains
\begin{equation}\label{ligh-cone-wave}
  4\frac{\partial^2\Psi}{\partial u\partial v}=-V_i(v-u)\Psi,
\end{equation}
then, the discretization scheme has the following form
\begin{eqnarray}
\Psi(N)&=&\Psi(W)+\Psi(E)-\Psi(S)\nonumber\\
&&-\frac{\Delta^2}{8}V(S)\left[\Psi(W)+\Psi(E)\right]+\mathcal{O}(\Delta^4),\label{ci1}
\end{eqnarray}
where the point $N$, $M$, $E$ and $S$ are the points of a rhombus of a grid with step $\Delta$ in the discretized $u$-$v$ plane, as follows: $S=(u,v)$, $W=(u+\Delta,v)$, $E=(u,v+\Delta)$ and $N=(u+\Delta,v+\Delta)$. The initial perturbation is encoded on the two axes $(u_0,v_0)$, which can be assumed as a Gaussian wave packet without loss of generality, since quasi-normal modes and the asymptotical behavior of perturbations do not depend on initial conditions (as confirmed by many numerical simulations). The temporal evolution of the perturbations is determined using the points of the diagonal.

\begin{figure}
\centering
\resizebox{\linewidth}{!}{\includegraphics*{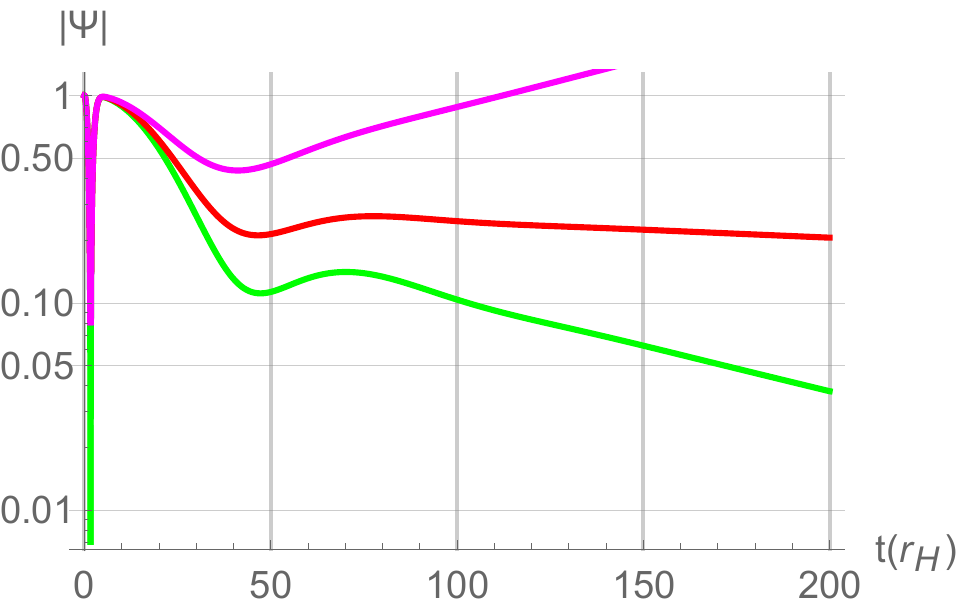}}
\caption{Temporal profiles for scalar-type perturbations for $r_H/r_C=0.75$, $\ell=2$ and $\alpha=0.118$, $0.121$, $0.126$ for green, red, and magenta profiles, respectively.}\label{profilescalar}
\end{figure}

To catch the threshold of (in)stability, we obtained a great number of time-domain profiles for various values of parameters. On Fig.~\ref{profilescalar} we present a typical behavior of the profiles, when crossing the threshold: as we approach to the instability region the purely damping mode first becomes dominant, and, finally, starts growing unboundedly, leading to the instability \cite{Cuyubamba:2016cug}.

\section{Parametric region of stability}
We obtain the threshold of instability for $\ell=2$ by analysing the temporal profiles and looking for an unboundedly growing perturbations. We have observed such behavior in the scalar sector only, while the perturbations of vector type remained stable.

\begin{figure*}
\centering
\resizebox{\linewidth}{!}{\includegraphics*{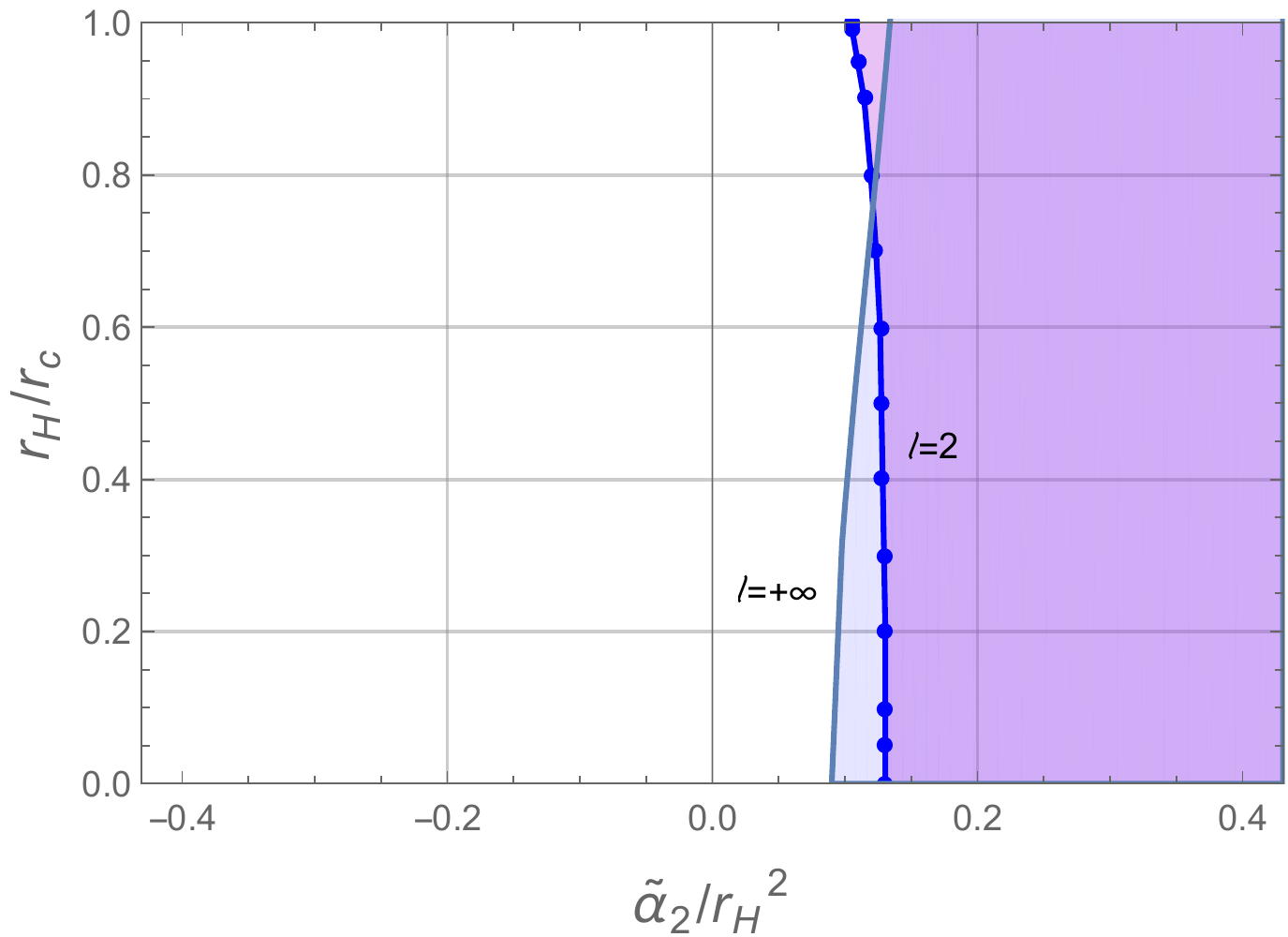},\includegraphics*{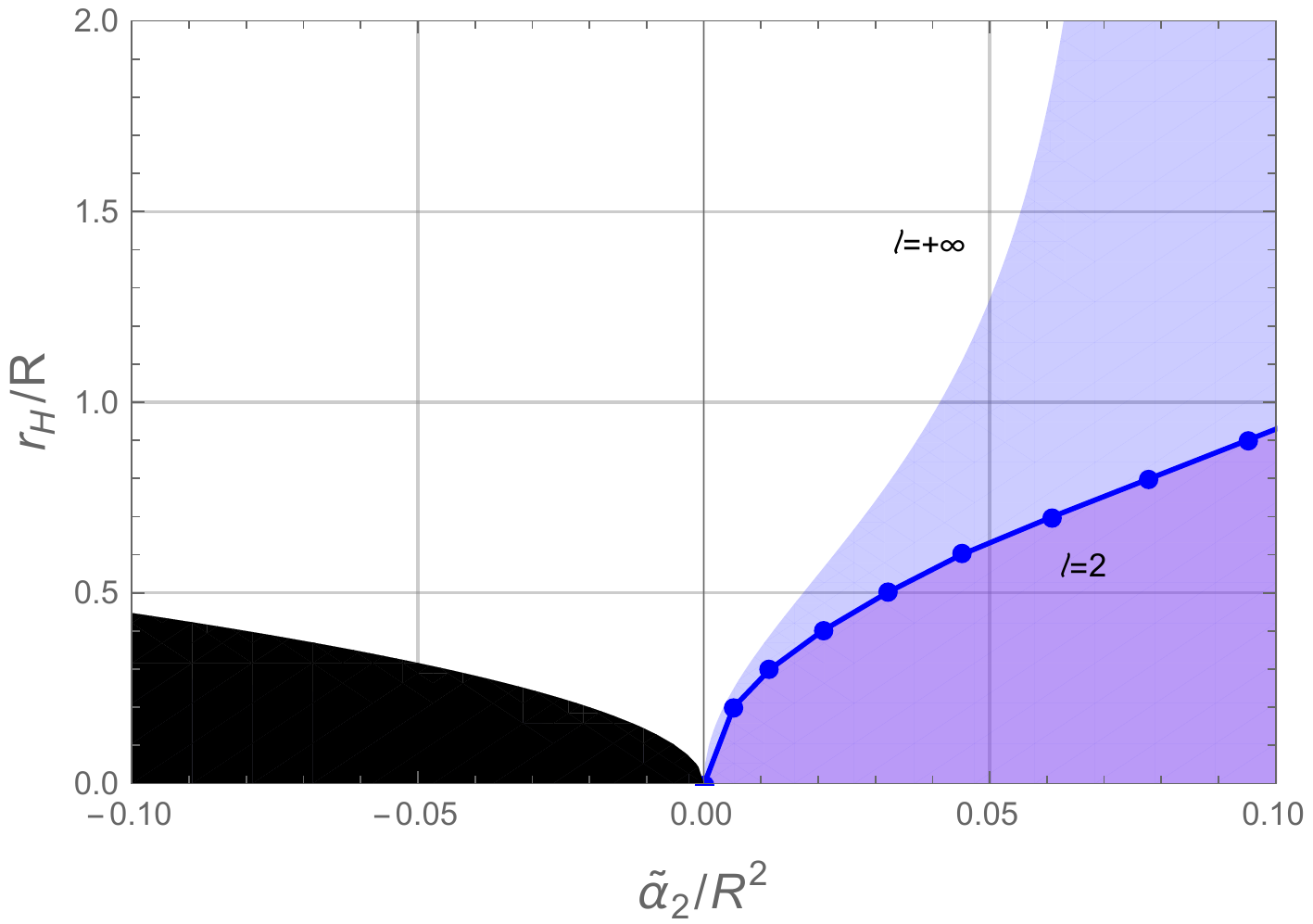}}
\caption{Complete parametric region of instability for 4-dimensional Gauss-Bonnet black hole asymptotically de Sitter (left) and Anti-de Sitter (right). The dotted blue points correspond to the $\Lambda$-instability and the soft curve is the eikonal instability.}\label{parametricregion}
\end{figure*}

\begin{figure}
\centering
\resizebox{\linewidth}{!}{\includegraphics*{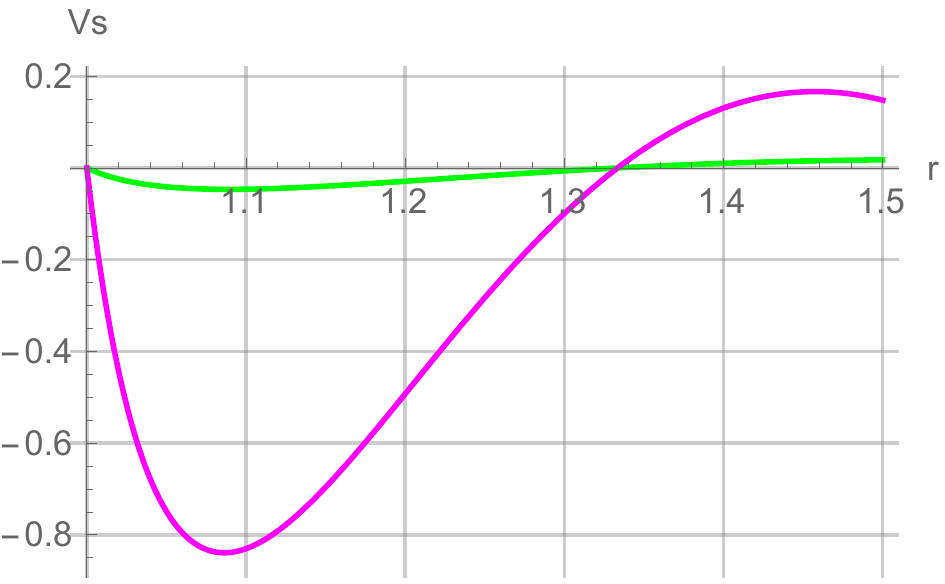}}
\caption{Effective potential to scalar-type gravitational perturbations for $r_H/r_C=75/100$, $\tilde{\alpha}_2=0.4r_H^2$ and $\ell=2$($\ell=10$) for green(magenta) curve.}\label{effectl}
\end{figure}

In the de Sitter spacetime ($\Lambda>0$), the overlap between the region of eikonal instability \cite{Konoplya:2020juj} and the $\Lambda$-instability (for $\ell=2$) is shown in the figure~\ref{parametricregion}. When the cosmological constant is small, black holes are unstable for sufficiently large value of $\tilde{\alpha}_2$-coupling parameter, and the threshold value is dominated by the eikonal instability, \ie there exists a finite critical value $\tilde{\alpha}_{crit}$ for a given cosmological constant $\Lambda$, such that for $\tilde{\alpha}_2>\tilde{\alpha}_{crit}$ black-hole perturbations are unstable at sufficiently large $\ell$. The reason for this phenomenon is that, as $\ell$ grows, the negative gap in the effective potential becomes deeper as shown on Fig.~\ref{effectl}. The increasing of the cosmological constant $\Lambda$ leads to increasing of $\tilde{\alpha}_{crit}$ until the perturbations of $\ell=2$ become unstable. One can see that, when $r_H/r_C\gtrapprox0.75$, the instability region is defined by $\ell=2$ ($\Lambda$ instability), so that at the threshold the perturbations, corresponding to higher values of the multipole number are more stable. A similar behavior was observed for the five-dimensional Einstein-Gauss-Bonnet black hole \cite{Cuyubamba:2016cug}.

\begin{table}
    \begin{tabular}{|c|c|c|c|}
    \hhline{|====|}
    \multicolumn{2}{|c|}{$\Lambda$-instability (dS)} & \multicolumn{2}{|c|}{$\Lambda$-instability (AdS)}\\
    \hline
    $r_H/r_c$ & $\tilde{\alpha}_2/r_H^2$ & $r_H/R$ & $\tilde{\alpha}_2/R^2$\\
    \hhline{|====|}
    0 & 0.131 & 0 & 0 \\
    0.05 & 0.131 & 0.2 & 0.0052 \\
    0.1 & 0.131 & 0.3 & 0.0116 \\
    0.2 & 0.131 & 0.4 & 0.0207 \\
    0.3 & 0.130 & 0.5 & 0.032 \\
    0.4 & 0.129 &  0.6 & 0.045 \\
    0.5 & 0.128 & 0.7 & 0.061 \\
    0.6 & 0.127 & 0.8 & 0.078\\
    0.7 & 0.124 & 0.9 & 0.095\\
    0.8 & 0.120 & 1 & 0.112\\
    0.9 & 0.115 & & \\
    0.95 & 0.110 & & \\
    0.999 & 0.105 & &\\
    \hhline{|====|}
    \end{tabular}
  \caption{Critical values of $\tilde{\alpha}_2$ corresponding to the $\Lambda$-instability for scalar-type gravitational perturbation}
  \label{tab:parametric2}
\end{table}

\begin{figure}
\centering
\resizebox{\linewidth}{!}{\includegraphics*{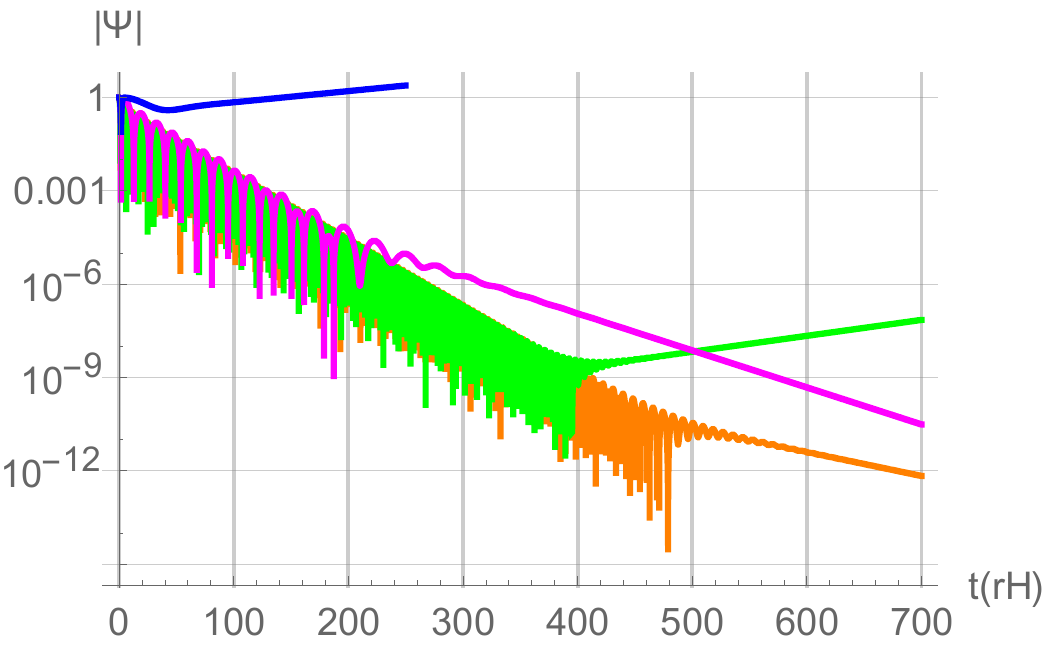}}
\caption{Temporal profiles for scalar perturbation close to the merge of eikonal and $\Lambda$-instability at $r_H/r_C=0.75$, $\tilde{\alpha}_2=0.125 r_H^2$. The blue, magenta, orange, and green profiles correspond to $\ell=2$, $4$, $12$ and $20$ respectively.}\label{Profileslgreatherthan2}
\end{figure}

As we can see in the left panel of Fig.~\ref{parametricregion}, close to $r_H/r_C\approx 0.75$, both threshold (of eikonal- and $\Lambda$-instability) curves cross. In this region, for $\tilde{\alpha}_2>\tilde{\alpha}_{crit}$  we observe both types of instability: the Fig. \ref{Profileslgreatherthan2} shows that, when $\tilde{\alpha}_2=0.125 r_H^2$, the time-domain profile for $\ell=2$ shows unbounded growth while the perturbations for $2<\ell<16$ are stable, and for larger values of $\ell$ they are unstable.

\begin{table*}
\begin{tabular}{|c|c|c|c|c|c|c|c|}
  \hhline{|========|}
   $r_H/r_C$ & $\tilde{\alpha}_2/r_H^2=-0.45$ & $\tilde{\alpha}_2/r_H^2=-0.3$ & $\tilde{\alpha}_2/r_H^2=-0.2$ & $\tilde{\alpha}_2/r_H^2=-0.1$ & $\tilde{\alpha}_2/r_H^2=0$ & $\tilde{\alpha}_2/r_H^2=+0.05$ & $\tilde{\alpha}_2/r_H^2=+0.08$\\
  \hhline{|========|}
   \multirow{2}{*}{$0$} & - - - - - - - & 1.3006-0.4643i & 1.0282-0.2615i & 0.8588-0.1987i & 0.7473-0.1779i & 0.7076-0.1724i & 0.6875-0.1689i \\
   & - - - - - - - & 1.9365-0.5454i & 1.3717-0.3608i & 1.0055-0.2486i & 0.7473-0.1779i & 0.6509-0.1675i & 0.6083-0.1690i \\
   \cline{1-8}
   \multirow{2}{*}{$0.1$} & - - - - - - - & 1.2933-0.4553i & 1.0159-0.2591i & 0.8446-0.1967i & 0.7313-0.1749i & 0.6904-0.1689i & 0.6695-0.1653i\\
   & - - - - - - - & - - - - - - - & 1.3519-0.3532i & 0.9888-0.2436i & 0.7313-0.1749i &  0.6340-0.1646i & 0.5906-0.1663i \\
   \cline{1-8}
   \multirow{2}{*}{$0.2$} & - - - - - - - & 1.2673-0.4319i & 0.9817-0.2519i & 0.8065-0.1907i & 0.6890-0.1668i & 0.6455-0.1597i & 0.6229-0.1556i \\
   & - - - - - - - & 1.8440-0.5055i & 1.2980-0.3329i & 0.9439-0.2305i & 0.6890-0.1668i & 0.5903-0.1568i & 0.5447-0.1586i\\
   \cline{1-8}
   \multirow{2}{*}{$0.3$} & 0.9713-0.9662i & 1.2200-0.3983i & 0.9282-0.2398i & 0.7495-0.1807i & 0.6279-0.1543i & 0.5816-0.1441i & 0.5573-0.1411i \\
   & - - - - - - - & 1.7407-0.4630i & 1.2166-0.3039i & 0.8768-0.2113i & 0.6279-0.1543i & 0.5281-0.1441i & 0.4801-0.1462i\\
   \cline{1-8}
   \multirow{2}{*}{$0.4$} & 0.9213-1.0283i & 1.1486-0.358i & 0.8563-0.2224i & 0.6769-0.1663i & 0.5535-0.1380i & 0.5056-0.1282i & 0.4802-0.1230i\\
   & - - - - - - - & 1.6067-0.4112i &1.1122-0.2690i & 0.7918-0.1879i & 0.5535-0.1380i & 0.4545-0.1280i & 0.4045-0.1294i\\
   \cline{1-8}
   \multirow{2}{*}{$0.5$} & 0.8067-1.1029i & 1.0516-0.3133i & 0.7664-0.1995i & 0.5912-0.1478i & 0.4702-0.1186i & 0.4227-0.1082i & 0.3973-0.1026i\\
   & - - - - - - - & 1.4444-0.3533i & 0.9871-0.2303i & 0.6920-0.1614i & 0.4702-0.1186i & 0.3750-0.1086i & 0.3243-0.1093i\\
   \cline{1-8}
   \multirow{2}{*}{$0.6$} & 2.2776-1.0892i$*$ & 0.9269-0.2649i & 0.6581-0.1741i & 0.4939-0.1251i & 0.3810-0.0975i & 0.3366-0.0867i & 0.3129-0.0811i\\
   & 3.2118-1.0711i & 1.2533-0.2914i & 0.8422-0.1893i & 0.5792-0.1326i & 0.3810-0.0975i & 0.2936-0.0873i & 0.2446-0.0870i\\
   \cline{1-8}
   \multirow{2}{*}{$0.7$} & 1.1928-0.8346i & 0.7706-0.2123i & 0.5304-0.1379i & 0.3860-0.0987i & 0.2881-0.0739i & 0.2499-0.0645i & 0.2296-0.0595i\\
   & 2.7662-0.8308i & 1.0293-0.2268i & 0.6761-0.1465i & 0.4541-0.1021i & 0.2881-0.0739i & 0.2135-0.0650i & 0.1691-0.0638i\\
   \cline{1-8}
   \multirow{2}{*}{$0.75$} & 1.8064-0.7059i & 0.6787-0.1840i & 0.4587-0.1192i & 0.3282-0.0842i &0.2408-0.0619i & 0.2069-0.0534i & 0.1889-0.0490i\\
   & 2.5036-0.7073i & 0.9023-0.1936i & 0.5842-0.1244i & 0.3870-0.0862i & 0.2408-0.0619i & 0.1746-0.0538i & 0.1340-0.0521i\\
   \cline{1-8}
   \multirow{2}{*}{$0.8$} & 1.6027-0.5772i & 0.5758-0.1538i & 0.3812-0.0988i & 0.2679-0.0689i & 0.1930-0.0497i & 0.1643-0.0424i & 0.1491-0.0386i\\
   & 2.2049-0.5816i & 0.7627-0.1594i & 0.4855-0.1017i & 0.3167-0.0700i & 0.1930-0.0497i & 0.1367-0.0427i & 0.1013-0.0407i\\
   \cline{1-8}
   \multirow{2}{*}{$0.9$} & 1.0593-0.3182i & 0.3287-0.0858i & 0.2066-0.0534i & 0.1394-0.0359i & 0.0967-0.0250i & 0.0807-0.0208i & 0.0724-0.0187i\\
   & 1.4346-0.3214i & 0.4338-0.0867i & 0.2639-0.0538i & 0.1660-0.0361i & 0.0967-0.0250i & 0.0652-0.0209i & 0.0441-0.0191i\\
    \hhline{|========|}
\end{tabular}
\caption{Fundamental quasinormal modes for gravitational perturbations ($\ell=2$) in the de Sitter spacetime. The first and second rows correspond, respectively, to the perturbations of vector-type and scalar-type. $*$When $\tilde{\alpha}_2=-0.45r_H^2$ and $r_H/r_C=0.6$, we observed two dominant modes in the spectrum of the vector(axial) channel, the first overtone has a little higher decay rate, $\omega=$ 0.7905-1.1893i. The dashed-lines correspond to the non-oscillatory behaviour.}
\label{ReviewBH2}
\end{table*}

In Table \ref{ReviewBH2} we present the dominant quasinormal modes of gravitational perturbations for both scalar and vector channels in the region of stability. The quasinormal modes were obtained directly by fitting the time-domain data by a superposition of damping exponents, using the Prony method. The quasinormal ringing for vector-type perturbations for $\tilde{\alpha}_2=-0.45r_H^2$ and $r_H/r_C=0.6$ has two dominant modes in the spectrum almost the same values of the damping rate Im($\omega)$. As the cosmological horizon $r_C$ decreases ($\Lambda$ grows), the other mode diminishes its decay rate faster and becomes dominant in the spectrum, so that we observe a discontinuity in the real part of the fundamental mode.

Due to a symmetry between the effective potentials of the Schwarzschild-de Sitter black hole in four dimensions both types of the gravitational perturbations (vector and scalar) have the same quasinormal spectrum \cite{Zhidenko:2003wq}. In Table \ref{ReviewBH2} we see that the isospectrality is broken for $4D$ regularized Einstein-Gauss-Bonnet black holes. For negative values of $\tilde{\alpha}_2$ the fundamental mode of vector-type perturbations dominates in the spectrum, while for positive coupling the scalar channel clearly dominates for sufficiently large $\Lambda$. When the cosmological constant is small the positive $\tilde{\alpha}_2$ within the stability interval changes significantly only the real part of the dominant mode, and the decay rate for both channels remains almost the same (within the fitting accuracy).

We have also tested stability of the asymptotically anti-de Sitter black holes. As one can see in the right panel of Fig.~\ref{parametricregion}, we did not find an indication of instability in any channel in the parametric region, where the black holes do not show the eikonal instability \cite{Konoplya:2020juj}. We conclude, therefore, that the anti-de Sitter black hole are stable for

\begin{equation}\label{stableds}
  -\frac{\mu^2}{2}<\frac{\tilde{\alpha}_2}{R^2}<\frac{\mu^2\left(A(\mu)-\sqrt{3}\left(1+\mu^2\right)\sqrt{B(\mu)}\right)}{2 C(\mu)}
\end{equation}
where
\begin{eqnarray}
 A(\mu)&=&1+6\sqrt{3}\mu^2\left(1+\mu^2\right)+9\mu^2\left(-1+\mu^2+3\mu^4\right)\nonumber\\
 B(\mu)&=&-3+2\sqrt{3}+9\mu^4\left(2+\left(9+6\sqrt{3}\right)\mu^4\right)\\
 C(\mu)&=&-1-18\mu^4+27\mu^8\nonumber
\end{eqnarray}
and $\mu=r_+/R$. Taking the limit $\mu\to +\infty$ on the right-hand side expression of the inequality (\ref{stableds}), one can obtain the maximum critical value $\tilde{\alpha}_{max}=\left(3-\sqrt{3+2\sqrt{3}}\right)R^2/6$. Thus, any $4D$-dimensional black hole asymptotically Anti-de Sitter in regularized Einstein-Gauss-Bonnet theory is unstable for $\tilde{\alpha}_2>\tilde{\alpha}_{max}$.

In Table~\ref{AdSBH3} we show the quasinormal frequencies for both channels of the gravitational perturbations of the Einstein-Gauss-Bonnet-AdS black hole. By considering the parameters $r_H$ and $\tilde{\alpha}_2$ within the stable region, we observe that the vector-type perturbations decay slower sector dominates for sufficiently large black holes, whereas for intermediate black holes the decaying rate lifetime of the scalar-type perturbations becomes longer.

The dashed lines, shown in Table~\ref{ReviewBH2} and Table~\ref{AdSBH3}, correspond to the non-oscillatory spectrum behaviour in the temporal profile, where the dominant mode is purely imaginary. Such non-oscillatory modes also appear for $\tilde{\alpha}_2\geq 0$ in the case of Anti-de Sitter black holes within the stable stability region.

\begin{table*}
\begin{tabular}{|c|c|c|c|c|}
  \hhline{|=====|}
   $r_H/R$ & $\tilde{\alpha}_2/R^2=-0.4$ & $\tilde{\alpha}_2/R^2=-0.3$ & $\tilde{\alpha}_2/R^2=-0.2$ & $\tilde{\alpha}_2/R^2=-0.1$ \\
  \hhline{|=====|}
   \multirow{2}{*}{$2$} & 2.2462-0.9390i & 2.0375-1.1855i & 1.6682-1.555i & - - - - - - -\\
   & 6.1553-1.8762i & 6.0458-2.2994i & 5.8776-2.8476i & - - - - - - - \\
   \cline{1-5}
   \multirow{2}{*}{$1.9$} & 2.3378-0.8733i & 2.1413-1.1147i & 1.8072-1.4754i & 0.7586-2.1144i \\
   & 5.9166-1.6979i & 5.7950-2.1061i & 5.6145-2.6330i & 5.3875-3.1195i \\
   \cline{1-5}
   \multirow{2}{*}{$1.8$} & 2.3981-0.7910i & 2.1798-1.0036i & 1.8809-1.3496i & 1.0911-1.9904i \\
   & 5.6784-1.4949i & 5.5335-1.8472i & 5.3579-2.3472i & 5.016-2.9243i$*$\\
   \cline{1-5}
   \multirow{2}{*}{$1.7$} & 2.4451-0.6980i & 2.2711-0.9229i & 1.9964-1.2591i & 1.3408-1.8781i\\
   & 5.4459-1.2764i & 5.3095-1.6484i & 5.1202-2.1291i & 4.7622-2.7214i\\
   \cline{1-5}
   \multirow{2}{*}{$1.6$} & 2.5156-0.6124i & 2.3258-0.8161i & 2.0752-1.1392i & 1.5268-1.7313i\\
   & 5.2478-1.0778i & 5.0816-1.4059i & 4.8889-1.8629i & 4.5520-2.4556i\\
   \cline{1-5}
   \multirow{2}{*}{$1.5$} & 2.5642-0.5111i & 2.4032-0.7192i & 2.1688-1.0321i & 1.6961-1.6022i\\
   & 5.0486-0.8545i & 4.8884-1.1895i & 4.6822-1.6277i & 4.3400-2.1978i\\
   \cline{1-5}
   \multirow{2}{*}{$1.4$} & 2.6445-0.4173i & 2.4845-0.6170i & 2.2619-0.9195i & 1.8488-1.4673i \\
   & 4.9013-0.6558i & 4.7222-0.9717i & 4.4998-1.3913i & 4.1513-1.9416i \\
   \cline{1-5}
   \multirow{2}{*}{$1.3$} & 2.7253-0.3166i & 2.5631-0.5055i & 2.3487-0.7957i & 1.9865-1.3176i \\
   & 4.7772-0.4548i & 4.5768-0.7456i & 4.3378-1.1427i & 3.9874-1.6710i \\
   \cline{1-5}
   \multirow{2}{*}{$1.2$} & 2.8336-0.2181i & 2.6527-0.3893i & 2.4403-0.6657i & 2.1111-1.1695i\\
   & 4.7104-0.2751i & 4.4672-0.5248i & 4.2062-0.8955i & 3.8442-1.4087i \\
   \cline{1-5}
   \multirow{2}{*}{$1.1$} & 2.9575-0.1217i & 2.7655-0.2722i & 2.5451-0.5329i & 2.2329-1.0092i  \\
   & 4.6793-0.1222i & 4.4101-0.3217i & 4.1172-0.6597i & 3.7305-1.1433i  \\
   \cline{1-5}
   \multirow{2}{*}{$1$} & 3.1033-0.0430i & 2.9080-0.1572i & 2.6628-0.3903i & 2.3576-0.8452i  \\
   & 4.6680-0.0288i & 4.4101-0.1487i & 4.0687-0.4276i & 3.6527-0.8905i \\
   \cline{1-5}
   \multirow{2}{*}{$0.9$} & - - - - - - - & 3.0777-0.0581i & 2.8085-0.2437i & 2.4890-0.6714i  \\
   & - - - - - - - & 4.4354-0.0356i & 4.6679-0.2181i & 3.6147-0.6427i  \\
   \cline{1-5}
   \hhline{|=====|}
\end{tabular}
\caption{Fundamental quasinormal modes for gravitational perturbations ($\ell=2$) in the Anti-de Sitter spacetime. The first and second rows correspond, respectively, to the perturbations of vector-type and scalar-type. The dashed-lines correspond to the non-oscillatory spectrum behaviour.}
\label{AdSBH3}
\end{table*}

\section{Conclusion}
While quasinormal modes and stability of the higher dimensional Einstein-Gauss-Bonnet \cite{Konoplya:2004xx,Abdalla:2005hu,Zhidenko:2008fp,Gonzalez:2017gwa}, or four dimensional Einstein-dilaton-Gauss-Bonnet black holes \cite{Blazquez-Salcedo:2020rhf,Blazquez-Salcedo:2016enn,Zinhailo:2019rwd,Konoplya:2019hml} are relatively well, though not completely, studied, the analysis of stability and spectra of the novel $4D$ Einstein-Gauss-Bonnet black holes is limited by only a few studies, where only the eikonal type of instability was discussed \cite{Konoplya:2020juj,Konoplya:2020bxa}.

Here we have obtained the full parametric region of stability for the four-dimensional Einstein-Gauss-Bonnet-(anti)de Sitter black holes. For the sufficiently large $\Lambda$, in addition to the eikonal instability, we have observed $\Lambda$-instability, expanding the instability region previously reported in \cite{Konoplya:2020bxa,Konoplya:2020juj}. This is similar to the instability observed for $5$-dimensional Einstein-Gauss-Bonnet-de Sitter black holes \cite{Cuyubamba:2016cug}, where the instability in scalar channel is obtained by overlapping of eikonal- and $\Lambda$- regions of instability. We did not find indications of $\Lambda$-instability for the asymptotically anti-de Sitter black holes, when they do not suffer from the eikonal instability. In addition, we have  observed that the isospectrality between vector- and scalar-type perturbations is broken for $4D$-regularized Einstein-Gauss-Bonnet black holes for any value of the $\Lambda$-term.

In order to obtain the complete picture of the stability region, it would be interesting to extend out our analysis to the $4D$ Einstein-Gauss-Bonnet black holes with electric charge.

\begin{acknowledgments}
I would like to thank Alexander~Zhidenko and Roman~Konoplya for useful discussions and critical reading of the manuscript. The author acknowledges the support by the Alexander von Humboldt Foundation, Germany.
\end{acknowledgments}

\end{document}